\documentclass[12pt]{iopart}

\usepackage{graphicx}
\usepackage{makecell}
\usepackage{hyperref}
\hypersetup{
  colorlinks=true,
  citecolor=blue,
  linkcolor=blue,
  urlcolor=blue,
  hypertexnames=false,
  pdftitle={Strain-controlled crystalline-amorphous transition and flat-band tuning in buckled silicon kagome},
  pdfauthor={Chenhaoyue Wang and Amartya S. Banerjee},
  pdfsubject={Strain-tunable structural metastability and kagome-derived flat bands in a two-dimensional silicon allotrope},
  pdfkeywords={silicon kagome lattice, flat bands, strain engineering, two-dimensional materials, crystalline-amorphous transition}
}

\usepackage{scalerel}
\usepackage{tikz}
\usetikzlibrary{svg.path}
\definecolor{orcidlogocol}{HTML}{A6CE39}
\tikzset{
  orcidlogo/.pic={
    \fill[orcidlogocol] svg{M256,128c0,70.7-57.3,128-128,128C57.3,256,0,198.7,0,128C0,57.3,57.3,0,128,0C198.7,0,256,57.3,256,128z};
    \fill[white] svg{M86.3,186.2H70.9V79.1h15.4v48.4V186.2z}
                 svg{M108.9,79.1h41.6c39.6,0,57,28.3,57,53.6c0,27.5-21.5,53.6-56.8,53.6h-41.8V79.1z M124.3,172.4h24.5c34.9,0,42.9-26.5,42.9-39.7c0-21.5-13.7-39.7-43.7-39.7h-23.7V172.4z}
                 svg{M88.7,56.8c0,5.5-4.5,10.1-10.1,10.1c-5.6,0-10.1-4.6-10.1-10.1c0-5.6,4.5-10.1,10.1-10.1C84.2,46.7,88.7,51.3,88.7,56.8z};
  }
}
\newcommand\orcidicon[1]{\href{https://orcid.org/#1}{\mbox{\scalerel*{
\begin{tikzpicture}[yscale=-1,transform shape]
\pic{orcidlogo};
\end{tikzpicture}
}{|}}}} 

\begin{document}

\title[Strain-controlled transition and flat-band tuning]{Strain-controlled crystalline--amorphous transition and flat-band tuning in buckled silicon kagome}

\author{Chenhaoyue Wang  \orcidicon{0000-0000-0000-0000}}
\address{Department of Materials Science and Engineering, University of California, Los Angeles, CA 90095, USA}
\author{Amartya S. Banerjee \orcidicon{0000-0001-5916-9167}}
\address{Department of Materials Science and Engineering, University of California, Los Angeles, CA 90095, USA}
\ead{asbanerjee@ucla.edu}

\begin{abstract}
Flat electronic bands in an elemental two-dimensional material would provide a chemically simple setting in which electron interactions compete with a suppressed kinetic-energy scale. Here we propose a buckled silicon kagome lattice (SiKL), an unfunctionalized six-atom monolayer composed of bond-linked Si$_3$ triangles and dodecagonal pores. The planar parent hosts a dispersionless Kohn--Sham band near the Fermi level but is unstable to out-of-plane distortions. Following three soft zone-centre phonons and relaxing the displaced structures yields two nearly degenerate buckled configurations with heights of 1.22 and 1.40~\AA. The high-buckling configuration retains a partially flat kagome-derived band near the Fermi level. Biaxial tension simultaneously controls its lattice dynamics and electronic dispersion: at 10\% strain, the calculated bandwidth reduces significantly, the associated density-of-states peak approaches the Fermi level, and the softest  phonon frequency hardens. In $6\times6$ ab initio molecular-dynamics simulations at 315~K, the unstrained network transitions to disorder, whereas the strained network remains ordered, indicating strain-induced finite-temperature metastability. Fifty-nanosecond classical molecular-dynamics simulations of $36\times36$ sheets further reveal a strain-controlled crystalline--amorphous transition and a local-bonding crossover near 2\% strain. Low-strain trajectories show gradual, two-stage disordering, whereas higher strains undergo an abrupt, first-order-like collapse, with the transition temperature reaching approximately 600~K at 10\% strain. An exploratory Ag(111) substrate model suggests that epitaxial mismatch could supply comparable tension, retain a narrow SiKL band, and preserve crystalline order well above room temperature. Unlike chemically passivated or hybrid-lattice silicon kagome proposals designed mainly as conventional semiconductors, SiKL is elemental and uses strain alone to couple thermal metastability, bond rearrangement, and near-Fermi flat-band tuning. Buckled SiKL is therefore a candidate material platform for strain-controlled flat-band and electronic correlation physics.
\end{abstract}

\section{Introduction}

Flat electronic bands provide a route to regimes in which the kinetic energy of carriers is strongly reduced relative to Coulomb, exchange, and spin--orbit energy scales. When the bandwidth $W$ is sufficiently small, interaction effects that are perturbative in ordinary dispersive bands can instead determine the ground state, enabling correlated insulating behaviour, itinerant magnetism, superconductivity, and topological phases \cite{leykam_artificial_2018,neupert_charge_2022,po_origin_2018,yankowitz_tuning_2019}. The correlated and superconducting states observed in magic-angle twisted bilayer graphene established the power of this principle \cite{cao_correlated_2018,lisi_observation_2021}. At the same time, the stringent twist-angle and stacking requirements of moir\'e materials motivate the search for atomic monolayers in which a narrow band is generated directly by local lattice topology.

The kagome lattice is a canonical setting for such physics. In the nearest-neighbour model, destructive interference on a network of triangular motifs creates compact localized states and a flat band, together with Dirac crossings and van Hove singularities \cite{Mielke_1991,zhou_s_2014,bergholtz_topological_2013,yin_topological_2022, sharma2025strain}. In real materials, further-neighbour hopping, orbital mixing, buckling, spin--orbit coupling, and substrate hybridization generally give the ideal flat band a finite dispersion. This departure from the ideal model makes the electronic bandwidth and band alignment potentially controllable by strain, chemistry, and interfaces. Related quasi-one-dimensional materials reinforce this design principle: specialized symmetry adapted first principles calculations \cite{ghosh2019symmetry, banerjee2021ab, yu2022density, agarwal2024solution, pathrudkar2022machine} show that carbon kagome nanotubes inherit near-Fermi flat bands from kagome graphene and exhibit deformation-induced electronic transitions, while phosphorus carbide nanotubes combine Dirac fermions and multiple flat bands with strain-driven structural and quantum phase transitions \cite{yu_carbon_2024,sharma_dirac_2026}. These examples motivate the broader question of whether mechanical deformation can likewise stabilize a pristine two-dimensional elemental kagome network while tuning its narrow-band electronic structure.  An atomically thin, single-element kagome network would be especially attractive because it could retain a comparatively transparent connection between geometry, bonding, and electronic interference.

Kagome-graphene networks based on bond-linked triangular units \cite{zhong2016coexistence, sarikavak2020structural} have been predicted to host flat-band ferromagnetism and Wigner crystallization \cite{chen_ferromagnetism_2018}. Silicon is a natural analogue to investigate because low-coordinated Si readily buckles and mixes nominal $sp^2$- and $sp^3$-like bonding, while epitaxial silicene phases have already been prepared on metallic substrates \cite{vogt_silicene_2012}. Silicon also offers an unusual bridge between emergent quantum materials and the mature infrastructure of semiconductor growth and processing. A strain-tunable flat-band allotrope made only of Si would therefore be interesting both as a fundamental quantum-material platform and as a possible route towards integrating narrow-band physics with established silicon technologies.

Several distinct silicon systems with kagome-related structures have previously been proposed or observed, and they provide context for the present work. Leenaerts \textit{et al.} studied three-dimensional group-14 kagome frameworks, including Si \cite{leenaerts_stable_2015}. Matusalem \textit{et al.} investigated a nonplanar two-dimensional group-14 ``basket-weave'' or kagome allotrope with a connectivity distinct from the linked-triangle network considered here \cite{matusalem_stability_2015}. An electronic kagome superlattice and flat band were observed in twisted multilayer silicene, but the kagome pattern there is an emergent electronic feature rather than the atomic lattice of a monolayer allotrope \cite{li_realization_2018}. Flat bands have also been observed in a distorted Ag kagome overlayer on Si(111); in that system the kagome sites and dominant orbitals are Ag rather than an elemental silicon allotrope \cite{lee_atomically_2024}. A kagome-like Si phase was assigned on Al(111) \cite{sassa_kagome-like_2020}, although a subsequent structural analysis favored an Al-embedded silicene model for that surface phase \cite{sato_surface_2025}. More recent theoretical designs have focused on hybrid honeycomb--kagome silicon \cite{sang_semiconducting_2021}, a semiconducting kagome-silicon network for field-effect transistors \cite{sang_geometric_2023}, and hydrogenated kagome silicene Si$_6$H$_6$ with strain- and substrate-tunable semiconducting and optical properties \cite{zhang_si6h6_2025}.

\begin{table}[t]
\caption{Representative silicon kagome-related systems and the features that distinguish them from the present work.}
\label{tab:prior}
\centering
\footnotesize
\renewcommand{\arraystretch}{1.20}
\begin{tabular}{|p{0.27\textwidth}|p{0.64\textwidth}|}
\hline
System & Primary distinction or design objective \\
\hline
Three-dimensional group-14 kagome frameworks \cite{leenaerts_stable_2015} & Bulk, fourfold-coordinated frameworks rather than a freestanding 2D monolayer. \\
\hline
Two-dimensional basket-weave group-14 allotropes \cite{matusalem_stability_2015} & Elemental and nonplanar, but with a different atomic connectivity and without the coupled strain--flat-band--amorphization analysis developed here. \\
\hline
Twisted multilayer silicene \cite{li_realization_2018} & An emergent electronic kagome superlattice generated by twisting, rather than an atomic kagome allotrope. \\
\hline
Ag kagome overlayer on Si(111) \cite{lee_atomically_2024} & Experimentally observed $d$-orbital flat bands of a distorted Ag kagome layer; the kagome sites are Ag rather than Si. \\
\hline
Hybrid and semiconducting kagome silicon \cite{sang_semiconducting_2021,sang_geometric_2023} & Hybrid-lattice engineering directed towards band-gap opening, carrier transport, and transistor applications. \\
\hline
Hydrogenated Si$_6$H$_6$ \cite{zhang_si6h6_2025} & Chemical passivation produces an indirect-gap semiconductor with strain-tunable optical and transport properties. \\
\hline
Buckled SiKL (this work) & Unfunctionalized six-atom linked-triangle network in which strain alone controls finite-temperature metastability, crystalline--amorphous behaviour, and a near-Fermi kagome-derived flat band. \\
\hline
\end{tabular}
\end{table}

Here we propose a buckled silicon kagome lattice (SiKL), an
unfunctionalized six-atom monolayer composed of bond-linked Si$_3$
triangles and dodecagonal pores. Its novelty lies in the conjunction of
four features: a specific kagome-graphene-derived topology; buckling
obtained by following the soft modes of the unstable planar parent;
retention of a narrow kagome-derived band near the Fermi level; and a
strain-only route that simultaneously enhances finite-temperature
persistence and tunes that band. Strain is already recognized as a
powerful control parameter in flat-band materials: heterostrain reshapes
moir\'e flat bands in twisted graphene, while periodic buckling and
folding can generate tunable flat bands and correlated states in
monolayer graphene
\cite{huder_heterostrain_2018,qiao_heterostrain_2018,
mao_buckled_2020,yang_origami_2022}. Strain has also been predicted
to shift kagome-derived flat bands and drive electronic or topological
transitions in chemically specific kagome compounds
\cite{kim_ni3sn_2024,xing_v3f8_2024}. The present work extends this
paradigm to a distinct regime: in an unfunctionalized elemental
two-dimensional lattice, biaxial tension not only narrows a
kagome-derived band near the Fermi level, but also suppresses the soft
structural instability, enhances finite-temperature persistence, and
modifies the crystalline-to-amorphous disordering pathway. Notably, as
discussed below, SiKL occupies a metastable branch of the silicon
allotrope landscape; its experimental realization will therefore likely
require kinetic trapping together with a substrate that provides both
geometric templating and tensile constraint. Existing literature on the demonstrated ability of epitaxial surfaces and interfacial alloys to select metastable low-dimensional silicon structures motivates the Ag(111)-based
proof of concept discussed here
\cite{vogt_silicene_2012,fleurence_epitaxial_2012,
kuchle_scaffold_2022}.

We investigate SiKL using density-functional theory (DFT), density-functional perturbation theory (DFPT), ab initio molecular dynamics (AIMD), and longer classical molecular dynamics (MD). We first map a soft-mode subspace of the planar lattice and identify two nearly degenerate buckled configurations. We then show that biaxial strain substantially narrows the kagome-derived band, suppresses the dominant soft-mode instability, and stabilizes the ordered network over the simulated finite-temperature trajectory. Large-scale MD reveals a non-monotonic strain dependence of the crystalline--amorphous transition temperature and a change from gradual, two-stage disordering to an abrupt, single-stage transition near a strain-induced bond-rearrangement crossover. Finally, we explore Ag(111) as a possible strain-imposing template for MBE or CVD growth. Throughout, we distinguish harmonic stability, finite-temperature metastability over a stated trajectory, and equilibrium thermodynamic stability. This hierarchy is particularly important for a soft two-dimensional lattice, for which a residual harmonic instability can coexist with substantially enhanced structural persistence at finite temperature.

\section{Computational methods}

\subsection{First-principles calculations}

First-principles calculations were performed with Quantum ESPRESSO \cite{giannozzi_quantum_2009}. The Perdew--Burke--Ernzerhof generalized-gradient approximation (PBE) \cite{perdew_generalized_1996} and the Perdew--Zunger local-density approximation (LDA) \cite{perdew_self-interaction_1981} were used to relax the planar silicon kagome lattice and compare its electronic structure. Periodic boundary conditions were applied in plane, with 20~\AA\ of vacuum normal to the monolayer. The PBE calculations used the PSLibrary PAW
pseudopotential, 
with nonlinear core correction, valence charge $Z_{\rm val}=4$, and valence configuration $3s^2 3p^2$. The LDA calculations also used
$Z_{\rm val}=4$. Self-consistent calculations used a 50~Ry wave-function cutoff, a
500~Ry charge-density cutoff, a $15\times15\times1$ Monkhorst--Pack mesh, Gaussian smearing of $10^{-3}$~Ry, and an electronic convergence threshold of $10^{-9}$~Ry. Structural relaxations were continued until the maximum force was below $5\times10^{-4}$~Ry/bohr and the in-plane stress components were below $10^{-2}$~kbar. Final band structures and densities of states were evaluated with a $45\times45\times1$ mesh and the 70~Ry wave-function cutoff and the 700~Ry charge-density cutoff.

Phonon dispersions were computed within DFPT using PBE \cite{baroni_phonons_2001}. Ground-state calculations used a $30\times30\times1$ electronic mesh, and dynamical matrices were sampled on a $5\times5\times1$ phonon wave-vector mesh. Biaxial strain is defined as $\varepsilon=(a-a_0)/a_0$, where $a_0$ is the relaxed unstrained in-plane lattice constant. Armchair and zigzag strains were applied along the corresponding principal directions.

\subsection{Ab initio molecular dynamics}

AIMD simulations were performed with SPARC using PBE \cite{ghosh_sparc_2017, ghosh_sparc_extended_2017} exchange correlation. Periodic boundary
conditions were used in plane and a Dirichlet boundary condition was used normal
to the sheet, with more than 20~\AA\ of vacuum. A real-space mesh spacing of
0.3~bohr was used. The SPARC AIMD runs used Gaussian electronic smearing and a
Nos\'e--Hoover canonical ensemble, 
with a 1~fs time step. No separate thermostat damping-time keyword was specified, so the
default SPARC Nos\'e--Hoover thermostat settings were used. The reported electronic and ionic convergence tolerances were $10^{-6}$ for the total energy and $5\times10^{-4}$ for forces. Both $3\times3$ and $6\times6$ supercells were examined at 315~K; the trajectory lengths associated with the figures are stated below.

\subsection{Classical molecular dynamics and transition metric}

Longer simulations were carried out with LAMMPS \cite{PLIMPTON19951}. 
Periodic boundary conditions were used in plane and a
nonperiodic boundary was used normal to the sheet. Si--Si
interactions in all reported 50~ns classical-MD transition-map simulations
were represented in LAMMPS using Tersoff potentials \cite{tersoff_empirical_1988}. 
The simulations used a Nos\'e--Hoover canonical ensemble, a 1~fs time step, and a
thermostat damping time of 0.1~ps. Initial velocities were sampled from a
Maxwell--Boltzmann distribution at the target temperature using one random seed
per strain--temperature condition. Production trajectories were run for up to
$5.0\times10^{7}$ steps, corresponding to 50~ns. Relaxed unit cells spanning
$0\leq\varepsilon\leq0.10$ in increments of 0.01 were replicated into
$36\times36\times1$ supercells for the strain--temperature calculations.

Structural order was monitored through radial distribution functions (RDFs), potential energy, and final atomic configurations. For each strain and observation time $t$, the selected crystalline RDF-peak intensity was fitted as a function of temperature with an error function. We define the crystalline--amorphous transition temperature $T_{\mathrm{CA}}(t)$ operationally as the midpoint of this fit. The principal map discussed below uses $t=50$~ns. Thus, $T_{\mathrm{CA}}$ is a finite-time, model-dependent transition metric rather than an equilibrium melting temperature; this definition allows meaningful comparison among the simulated strains while keeping the kinetic character of the calculation explicit.

For Ag-supported simulations, Ag--Ag interactions were described by an embedded-atom-method potential \cite{Foiles1986}. Si--Ag interactions were represented by the 12--6 Lennard--Jones form
\begin{equation}
\phi(r)=4\epsilon\left[\left(\frac{\sigma}{r}\right)^{12}-\left(\frac{\sigma}{r}\right)^6\right],
\end{equation}
with a 10~\AA\ cutoff. Lorentz--Berthelot mixing based on the tabulated parameters \cite{rappe_uff_1992,timpel1997silver} gave $\epsilon=0.0774$~eV and $\sigma=3.4175$~\AA\ for Si--Ag.

\section{Results and discussion}

\subsection{Planar silicon kagome lattice and soft-mode buckling}

\begin{figure}
\centering
\includegraphics[width=0.80\textwidth]{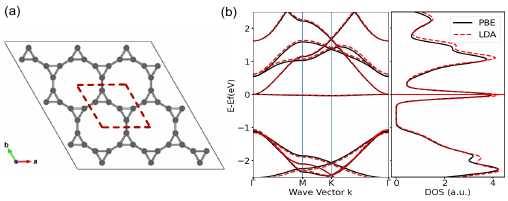}
\caption{(a) Top view of the planar SiKL parent lattice. The dashed polygon marks the primitive cell. (b) Kohn--Sham band structure (left) and density of states (right) calculated with LDA (red dashed curves) and PBE (black solid curves). Energies are referenced to the Fermi level.}
\label{fig:planar}
\end{figure}

The planar parent structure contains six Si atoms per primitive cell. Two Si$_3$ triangular units are connected by slightly longer inter-triangle bonds, generating a periodic array of dodecagonal pores (figure~\ref{fig:planar}). Relaxations with both PBE and LDA preserve the imposed planar symmetry. The inter-triangle bonds are approximately 0.05~\AA\ longer than the bonds within each triangle; the corresponding lattice parameters are listed in \ref{app:group14}. This linked-triangle topology is the direct silicon analogue of the kagome-graphene network and is distinct from the basket-weave and hybrid honeycomb--kagome silicon structures summarized in table~\ref{tab:prior}.

\begin{figure}
\centering
\includegraphics[width=1.0\textwidth]{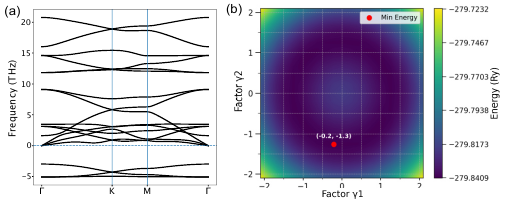}
\caption{(a) PBE phonon dispersion of planar SiKL. Negative plotted frequencies denote imaginary modes. (b) Representative relaxed-energy landscape for a fixed $\Gamma_3=0$ slice as a function of the $\Gamma_1$ and $\Gamma_2$ mode amplitudes. The red point marks the lowest-energy structure within the displayed slice.}
\label{fig:search}
\end{figure}

The planar lattice displays an essentially dispersionless band near the Fermi level along the plotted high-symmetry path, accompanied by a sharp density-of-states feature (figure~\ref{fig:planar}b). These signatures are characteristic of destructive interference in the linked-triangle geometry. The planar geometry is not, however, dynamically stable. Its PBE phonon dispersion contains several imaginary branches, including three unstable out-of-plane optical eigenvectors at the zone centre $\Gamma$ (figure~\ref{fig:search}a). We denote the degenerate pair by $\Gamma_1$ and $\Gamma_2$ and the third mode by $\Gamma_3$.

To identify the buckling patterns naturally selected by these soft modes, the planar coordinates were displaced along the three eigenvectors. For each chosen amplitude of $\Gamma_3$, the amplitudes of $\Gamma_1$ and $\Gamma_2$ were scanned from $-2$ to 2, after which every trial configuration was fully relaxed. Figure~\ref{fig:search}b shows one representative two-dimensional slice of this energy landscape. This soft-mode search is deliberately targeted: it explores the physically relevant neighbourhood of the planar saddle point rather than attempting a global search over all possible silicon networks.

\begin{figure}
\centering
\includegraphics[width=0.80\textwidth]{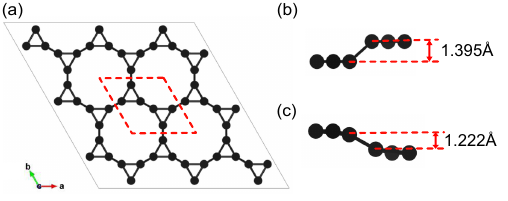}
\caption{Relaxed buckled SiKL structures: (a) top view of a $3\times3$ supercell and side views of (b) the HB and (c) the LB primitive cells. The indicated buckling height is the vertical separation between neighbouring triangular units.}
\label{fig:buckled}
\end{figure}

The search yields two buckled minima within the sampled soft-mode landscape, with buckling heights of 1.395 and 1.222~\AA\ (figure~\ref{fig:buckled}). We refer to them as the high-buckling (HB) and low-buckling (LB) structures. In both cases, the atoms within each Si$_3$ triangle remain nearly coplanar, while neighbouring triangles are vertically displaced relative to one another. The two structures differ in total energy by only 0.00025~eV per six-atom cell, or approximately 0.04~meV per atom. They should therefore be regarded as effectively degenerate at the numerical resolution of the present calculations. We use HB as the representative structure below because it is marginally lower in the reported energy and displays the larger geometric response to strain, not because the calculations uniquely establish it as the ground-state buckling pattern.

Table~\ref{tab:cohesive} compares cohesive energies recalculated with the same PBE workflow. HB and LB are approximately 1.14~eV per atom less bound than diamond Si, about 0.50~eV per atom less bound than silicene, and about 0.21~eV per atom less bound than the cited hybrid honeycomb--kagome allotrope. The proposed network therefore occupies a relatively high-energy, metastable branch of the silicon allotrope landscape. This energetic cost makes kinetic trapping, epitaxial constraint, or other nonequilibrium growth conditions central to its possible realization, but it does not by itself rule out synthesis: silicene and other low-dimensional allotropes illustrate how substrates and kinetic barriers can stabilize structures that are not the bulk thermodynamic ground state \cite{vogt_silicene_2012}.

\begin{table}[h]
\caption{PBE cohesive energies calculated with the workflow used in this study. More negative values indicate stronger binding.}
\label{tab:cohesive}
\centering
\footnotesize
\renewcommand{\arraystretch}{1.25}
\begin{tabular}{|l|c|}
\hline
Structure & Cohesive energy (eV atom$^{-1}$) \\
\hline
HB SiKL (this work) & $-4.27676$ \\
LB SiKL (this work) & $-4.27672$ \\
Silicene \cite{peng_mechanical_2013} & $-4.77621$ \\
Hybrid honeycomb--kagome Si \cite{sang_semiconducting_2021} & $-4.48198$ \\
Amorphous silicene \cite{gurbuz_systematic_2023} & $-4.73285$ \\
Diamond Si \cite{wyckoff_crystal_1963} & $-5.41730$ \\
Amorphous Si \cite{stich_amorphous_1991} & $-5.034$ \\
\hline
\end{tabular}
\end{table}

The resulting buckling is consistent with the tendency of low-coordinated silicon networks to mix nominal $sp^2$- and $sp^3$-like bonding. More importantly for the present study, it introduces an internal geometric degree of freedom through which in-plane strain can continuously modify Si--Si bond lengths, bond angles, hopping amplitudes, and the out-of-plane corrugation. This coupling makes strain a natural candidate for controlling both structural persistence and the kagome-derived electronic band.

\subsection{Strain-induced finite-temperature metastability}

\begin{figure}
\centering
\includegraphics[width=1.0\textwidth]{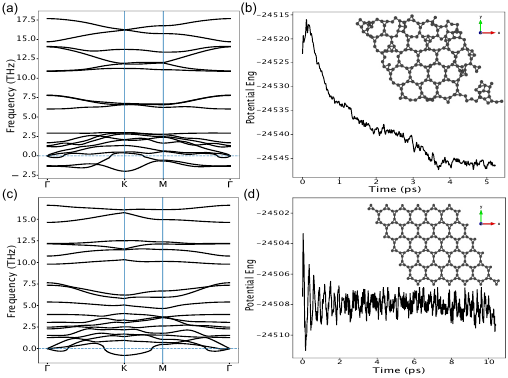}
\caption{PBE phonon dispersions and AIMD potential-energy histories for HB SiKL at 315~K: (a,b) unstrained and (c,d) under 10\% biaxial tensile strain. The AIMD cells contain $6\times6$ primitive cells. Insets show configurations averaged over the final 2~ps of the respective trajectories. Strain strongly hardens the soft modes and preserves the ordered linked-triangle network over the 10~ps strained trajectory.}
\label{fig:phonon_aimd}
\end{figure}

The HB structure remains soft in the harmonic approximation. Without strain, its phonon dispersion contains two imaginary branches and reaches a minimum plotted frequency of approximately $-1.98$~THz (figure~\ref{fig:phonon_aimd}a). The finite-temperature response is also strongly size dependent. A $3\times3$ AIMD supercell retains the ordered linked-triangle network over the available trajectory at 315~K, whereas a $6\times6$ cell develops extended disorder and pore disruption (figure~\ref{fig:phonon_aimd}b). Because the calculation is periodic in plane, this difference is best interpreted as the appearance of longer-wavelength distortions in the larger supercell rather than as a literal boundary effect.

Applying 10\% biaxial tensile strain strongly suppresses this instability.
The minimum imaginary frequency is hardened from approximately
$-1.98$ to $-0.64$ THz (figure~\ref{fig:phonon_aimd}c). Residual imaginary frequencies of this magnitude should be interpreted cautiously in soft two-dimensional systems, since low-frequency branches can be sensitive to phonon
interpolation, sum-rule enforcement, and anharmonic renormalization
\cite{petretto_convergence_2018,lin_invariance_2022,
pallikara_physical_2022}. Thus, the strained sheet may not be strictly harmonically stable at 0 K, but it is driven close to the
threshold of harmonic stability. Consistent with this interpretation,
the corresponding $6\times6$ AIMD trajectory at 315~K fluctuates around
a stationary mean energy and preserves the kagome-like network over
10~ps (figure~\ref{fig:phonon_aimd}d). Taken together, the phonon and
AIMD results show that biaxial strain converts the highly labile
unstrained sheet into a thermally metastable ordered structure, at least on the
simulated time and length scales. This is a key result of this work: strain alone,
without hydrogenation or substitutional chemistry, substantially
suppresses the dominant structural instability of the pristine SiKL
network.

\subsection{Flat-band tuning by strain}

\begin{figure}
\centering
\includegraphics[width=1.0\textwidth]{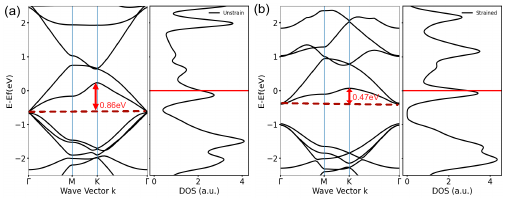}
\caption{Kohn--Sham band structures and densities of states of HB SiKL (a) without strain and (b) under 10\% biaxial tensile strain. The solid red line denotes the Fermi level in the DOS panel. The dashed red line and arrow indicate the energy interval used to quote the kagome-derived bandwidth. Biaxial strain reduces the displayed bandwidth from 0.86 to 0.47~eV.}
\label{fig:bands}
\end{figure}

Buckling broadens the idealized flat band of the planar parent but does not remove its kagome-derived character. In relaxed HB SiKL, the relevant band remains weakly dispersive near the Fermi level, particularly between $\Gamma$ and $M$, and has a reported bandwidth of 0.86~eV along the sampled path (figure~\ref{fig:bands}a). The accompanying density-of-states enhancement is consistent with survival of the linked-triangle interference after out-of-plane distortion, while the finite dispersion reflects additional hopping channels introduced by realistic Si bonding.

We relaxed HB SiKL under armchair, zigzag, and biaxial strains from $-0.10$ to 0.40 in increments of 0.01. Under increasing tensile strain, the Si--Si bonds elongate and the buckling height decreases. The flat-band width initially narrows, reaches a minimum near 10\% biaxial strain, and then broadens at larger deformation. The source calculations identify structural failure near strains of 0.23 in each uniaxial direction and 0.19 under biaxial loading. A complete mechanical characterization will require the corresponding stress--strain and strain-energy curves, but the present data already establish a broad interval over which the electronic structure can be tuned continuously before failure.

At 10\% biaxial strain, the reported bandwidth decreases from 0.86 to 0.47~eV (figure~\ref{fig:bands}b), a reduction of approximately 45\%. The associated density-of-states peak sharpens and moves closer to the Fermi level. This combination is scientifically significant even though the band is not mathematically dispersionless: correlation physics is controlled by the ratio of interaction scales to $W$, not by an absolute semantic threshold for the word ``flat''. The strain-induced reduction of $W$, together with the enhanced near-Fermi density of states, makes spin polarization, interaction-driven magnetism, superconductivity, and topological phases compelling targets for subsequent calculations at appropriate filling. Demonstrating any specific correlated phase will require spin-polarized and many-body analysis, but the present DFT result establishes a clear and unusually direct mechanical route for tuning the relevant one-electron energy scale.

\begin{figure}
\centering
\includegraphics[width=1.0\textwidth]{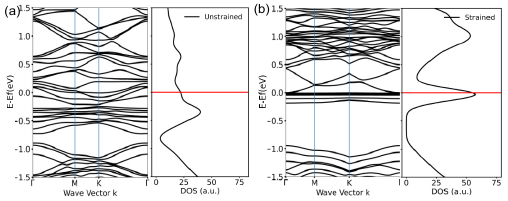}
\caption{Folded $6\times6$-supercell band structures (left) and densities of states (right) calculated from coordinates averaged over the final 3~ps of the 315~K AIMD trajectories: (a) unstrained and (b) under 10\% biaxial tension. The red line marks the Fermi level. The strained structure retains a pronounced near-Fermi DOS feature together with the ordered network.}
\label{fig:supercell_bands}
\end{figure}

The same structural--electronic contrast persists in the larger AIMD cells. Figure~\ref{fig:supercell_bands} compares electronic structures calculated from coordinates averaged over the final 3~ps of the $6\times6$ trajectories. The disordered unstrained configuration produces irregular bands and a broadened DOS, whereas the strained configuration retains a pronounced peak near the Fermi level and a set of weakly dispersive folded branches. These calculations support the persistence of the narrow-band signature in the ordered strained network. Because they use time-averaged coordinates and a folded supercell Brillouin zone, however, they should be interpreted qualitatively; snapshot-averaged spectral functions and band unfolding would provide the appropriate quantitative finite-temperature test.

\subsection{Strain-controlled crystalline--amorphous transition}

\begin{figure}
\centering
\includegraphics[width=1.0\textwidth]{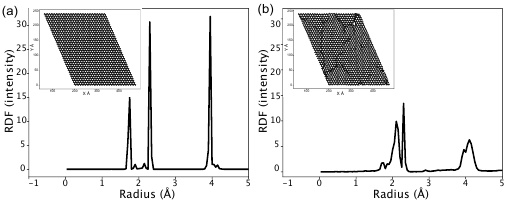}
\caption{RDFs and final top-view configurations of unstrained $36\times36$ HB sheets after 50~ns at (a) 90~K and (b) 110~K. The sharp low-temperature peaks correspond to the two characteristic Si--Si bond populations of crystalline SiKL; their broadening accompanies loss of the periodic pore network.}
\label{fig:rdf}
\end{figure}

The AIMD results establish the qualitative stabilizing effect of strain but are limited to picosecond trajectories and modest cell sizes. We therefore used the classical model to survey $36\times36$ sheets for up to 50~ns over a grid of strain and temperature. The RDF provides a direct measure of the persistence of the two principal Si--Si bond populations. At low temperature, sharp peaks and a periodic atomic configuration indicate a crystalline linked-triangle network (figure~\ref{fig:rdf}a). At higher temperature, the peaks broaden and lose intensity while the long-range pore pattern disappears, indicating transformation to an amorphous-like silicon network (figure~\ref{fig:rdf}b).

\begin{figure}
\centering
\includegraphics[width=1.0\textwidth]{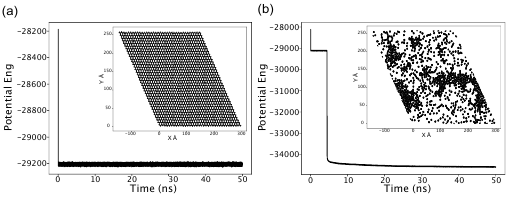}
\caption{Representative 50~ns classical-MD trajectories for
$\varepsilon=0.07$ HB SiKL. (a) At $T=460$~K, the crystalline SiKL
network is retained and the potential energy remains stationary after
initial equilibration. (b) At $T=530$~K, the sheet undergoes an abrupt
transformation to an amorphous-like network, accompanied by a sharp
drop in potential energy.}
\label{fig:md_energy}
\end{figure}

Representative energy histories illustrate the kinetic character of this transformation (figure~\ref{fig:md_energy}). In the ordered regime, the potential energy rapidly equilibrates and then fluctuates around a stationary value. In the disordering regime, a sudden energy decrease accompanies large-scale rearrangement of the network. The decrease does not represent ordinary melting of the bulk ground state; rather, it reflects relaxation of a strained, high-energy metastable sheet into a lower-energy amorphous-like configuration. The abrupt event is therefore appropriately described as first-order-like within the trajectory, while a rigorous equilibrium transition order would require free energies, size scaling, hysteresis, and independent seeds.

\begin{figure}
\centering
\includegraphics[width=1.0\textwidth]{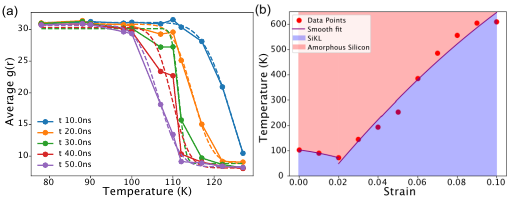}
\caption{(a) Selected RDF-peak intensity of unstrained HB SiKL versus temperature at several observation times; dashed curves are error-function fits. (b) Strain--temperature crystalline--amorphous transition map constructed from the 50~ns fit midpoints. Purple and red shading denote crystalline and amorphous-like outcomes, respectively, within the finite-time classical model.}
\label{fig:transition_map}
\end{figure}

For each strain, the intensity of the selected crystalline RDF peak decreases sigmoidally with temperature. Error-function fits at observation times from 10 to 50~ns are shown for the unstrained sheet in figure~\ref{fig:transition_map}a. The fitted midpoint shifts with observation time, as expected for an activated transformation. We use the 50~ns midpoint as $T_{\mathrm{CA}}$ and assemble the strain--temperature transition map in figure~\ref{fig:transition_map}b. The map is strongly non-monotonic. Small tensile strains initially lower $T_{\mathrm{CA}}$, but the trend reverses near 2\% strain. Above this crossover, the transition temperature rises rapidly, exceeds room temperature near 6\% strain, and reaches approximately 600~K at 10\% strain within the adopted potential.

\begin{figure}
\centering
\includegraphics[width=1.0\textwidth]{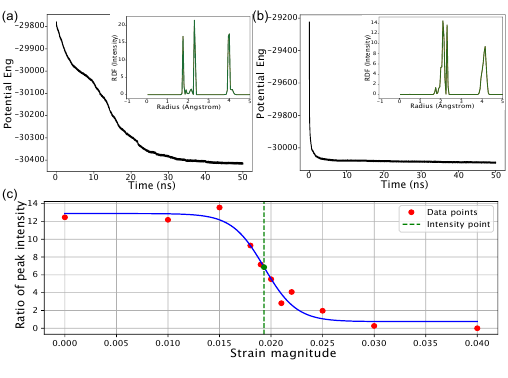}
\caption{Potential-energy histories over 50~ns at 120~K for (a) 1\% and (b) 3\% biaxial strain. Insets show RDFs at 50~ns for structures initially equilibrated at 80~K. (c) Ratio of the RDF-peak intensities near 1.8 and 2.1~\AA\ versus strain at 70~K. The dashed vertical line marks the fitted inflection at $\varepsilon=0.0196$, corresponding physically to a local-bonding crossover near 2\%.}
\label{fig:crossover}
\end{figure}

The non-monotonic transition line is linked to a strain-induced reorganization of the local bonding environment. At 70~K, the ratio of RDF-peak intensities near 1.8 and 2.1~\AA\ changes sigmoidally with strain (figure~\ref{fig:crossover}c). The fitted inflection occurs at $\varepsilon=0.0196$. We retain this value as the numerical fit result, while referring physically to a crossover near 2\% because the precision is limited by the strain sampling and potential model. Across this narrow range, weight is transferred between the intra-triangle and inter-triangle bond populations without immediate loss of long-range crystalline order. The resulting local rearrangement provides a microscopic explanation for the change in the thermal response: strain first perturbs the inequivalent bonds of the buckled network and then, beyond the crossover, produces a reorganized short-range environment that resists gradual disorder accumulation and raises the barrier to global amorphization.

\begin{figure}
\centering
\includegraphics[width=1.0\textwidth]{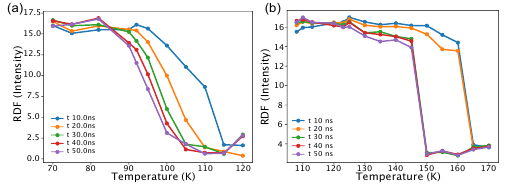}
\caption{Temperature dependence of the selected RDF-peak intensity at observation times from 10 to 50~ns for HB SiKL under (a) 1\% and (b) 3\% biaxial strain. The low-strain sheet loses order gradually over a broad interval, whereas the high-strain sheet shows a sharper transition.}
\label{fig:pathways}
\end{figure}

The low- and high-strain trajectories exhibit correspondingly different transition pathways. At 1\% strain, the potential energy evolves continuously over much of the trajectory before a final abrupt drop, and the RDF order parameter decreases over a broad temperature interval (figures~\ref{fig:crossover}a and \ref{fig:pathways}a). We interpret this as a two-stage process: a continuous, second-order-like regime of defect and bond-distortion accumulation followed by a first-order-like collapse once a critical degree of disorder is reached. At 3\% strain, the locally reorganized structure remains in a well-defined crystalline basin and then switches abruptly to the amorphous-like state, producing a narrower transition interval and a single-step, first-order-like event (figures~\ref{fig:crossover}b and \ref{fig:pathways}b). The qualifiers ``-like'' emphasize that the classification describes the kinetics and order-parameter signatures of the simulated trajectories rather than a thermodynamic proof of transition order. Even with that qualification, the change in pathway is a substantive finding: strain controls not only the temperature of amorphization but also how disorder nucleates and propagates through the SiKL network.

The resulting picture is therefore richer than a simple monotonic strengthening under tension. Very small strain initially destabilizes the inequivalent-bond network, but sufficient strain drives a local bond-population crossover that reorganizes the crystalline state, suppresses gradual disorder accumulation, and dramatically raises its finite-time transition temperature. This coupling among bond topology, transition pathway, and electronic bandwidth is one of the principal scientific merits of buckled SiKL.

\subsection{Ag(111) as a strain-imposing template and synthesis outlook}

The favourable structural and electronic regime occurs under biaxial tension, suggesting that epitaxial mismatch could provide a practical route to the required strain during growth. Ag(111) is a natural first substrate to examine because silver surfaces have played a central role in silicene synthesis and modelling \cite{vogt_silicene_2012,galashev_dft_2021}. We therefore constructed a proof-of-concept model containing a four-layer Ag(111) slab and a commensurate SiKL overlayer. The Si atoms and upper three Ag layers were relaxed, while the bottom Ag layer was fixed; the initial Si--Ag separation was 2.25~\AA. The selected coincidence geometry imposes approximately 11\% tensile strain on the silicon network, close to the regime in which the free-standing calculations show both strong flat-band narrowing and enhanced transition temperature.

\begin{figure}
\centering
\includegraphics[width=1.0\textwidth]{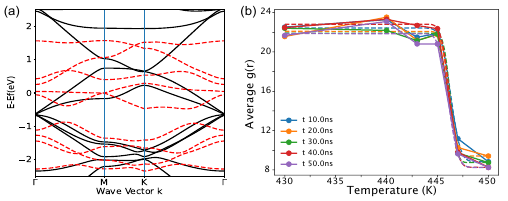}
\caption{(a) Band structures of the free-standing unstrained HB-SiKL reference
(black solid curves) and the free-standing peeled-off SiKL geometry obtained
after relaxation on Ag (red dashed curves). The red curve includes the
spontaneous strain generated during Ag-supported relaxation. (b) Selected crystalline RDF-peak amplitude versus temperature for the Ag-related classical-MD model at observation times from 10 to 50~ns.}
\label{fig:ag}
\end{figure}

Figure~\ref{fig:ag} compares the free-standing unstrained HB-SiKL reference with the
free-standing band structure of the SiKL layer extracted after relaxation on
Ag. Although the Ag-supported relaxation was initialized from the unstrained
HB geometry, the relaxed and peeled-off SiKL layer retains the spontaneous
strain generated during substrate relaxation. The kagome-derived band is reported to narrow further, to approximately 0.2~eV. This result provides a plausible link between the strain engineering identified above and an experimentally accessible epitaxial constraint. For a quantitative supported electronic structure, Si-projected spectral weight and band unfolding will be needed to separate the SiKL feature from metallic Ag bands and supercell folding.

The corresponding large-scale classical model contains approximately $5\times10^4$ atoms. Its RDF-fit midpoint is about 446~K at 50~ns (figure~\ref{fig:ag}b), showing that the mismatch-induced tensile geometry remains ordered to a temperature well above room temperature within this model. While these calculations do not by themselves confirm molecular beam epitaxy (MBE) or chemical vapor deposition (CVD) growth and are not intended to identify Ag(111) as a uniquely optimal substrate, the calculations above do establish a first example of a broader topology-selective growth strategy for SiKL. Experiments on low-dimensional silicon demonstrate that epitaxial surfaces can select metastable structures: silicene-related sheets have been reported on
Ag(111), epitaxial silicene forms through surface segregation on
ZrB$_2$(0001), anisotropic Ag(110) directs Si into ordered
nanoribbons, and an intervening Si--Ag surface alloy can scaffold the
silicene/Ag(111) interface
\cite{vogt_silicene_2012,fleurence_epitaxial_2012,
prevot_nanoribbons_2016,kuchle_scaffold_2022}. More broadly,
metastable silicon networks have been retained through precursor
removal and stress confinement, as demonstrated for open-framework
Si$_{24}$ and diamond-hexagonal Si nanoribbons
\cite{kim_openframework_2015,qiu_hexagonal_2015}. These precedents
suggest kinetically controlled MBE, potentially combined with a
reconstructed, alloyed, or patterned template, as the most direct
experimental route to buckled SiKL. Precursor-assisted CVD could be
explored as a complementary route if suitable precursor and surface
chemistry can be identified. For SiKL, however, the template must do
more than impose tensile mismatch: its adsorption-site registry must
preferentially nucleate the linked Si$_3$-triangle/dodecagon network
over silicene, compact Si clusters, and bulk-like phases, while
avoiding hybridization strong enough to destroy the near-Fermi flat
band. Future substrate screening should therefore compare coincidence
supercells, registry-dependent adhesion, Si diffusion and nucleation
barriers, surface reconstruction or alloying, and Si-projected
unfolded band structures, since balancing structural stabilization
against electronic preservation is a central substrate-design
criterion for two-dimensional silicon
\cite{chen_substrates_2016}.

\section{Conclusions}

We have identified a potentially novel, unfunctionalized silicon kagome allotrope in which strain controls structural metastability and a near-Fermi flat band through the same geometric degree of freedom. The planar six-atom linked-triangle lattice possesses an idealized kagome-derived flat band but is unstable to out-of-plane soft modes. Following those modes yields two nearly degenerate buckled structures with buckling heights of 1.22 and 1.40~\AA. The representative high buckling structure retains the electronic signature of the planar kagome network while acquiring the buckling flexibility characteristic of low-dimensional silicon.

Biaxial tension produces a coupled structural--electronic response. At 10\% strain, the calculated kagome-derived bandwidth is reduced from 0.86 to 0.47~eV, the near-Fermi DOS feature is enhanced, and the dominant soft phonon is strongly hardened. The strained $6\times6$ sheet remains ordered over a 10~ps AIMD trajectory at 315~K, whereas the unstrained sheet disorders. Thus, although a residual harmonic soft mode persists, strain alone creates clear finite-temperature metastability without chemical passivation or hybrid-lattice redesign.

The long classical trajectories further reveal a strain-controlled crystalline--amorphous transition. A local bond-population crossover occurs near 2\% strain, with a fitted inflection at 0.0196. Below this crossover, the network undergoes gradual, two-stage disorder accumulation followed by collapse; above it, the transition becomes abrupt and first-order-like. Within the adopted potential and a 50~ns window, the transition temperature rises to approximately 600~K at 10\% strain. The Ag(111) model suggests that epitaxial mismatch may supply a comparable tensile strain, preserve order to approximately 446~K in the classical interface model, and narrow the extracted SiKL band to about 0.2~eV.

The broader significance is that the present design does not obtain stability by converting the silicon kagome network into a chemically passivated or conventional semiconducting material. Instead, it preserves an elemental near-Fermi narrow band and uses a purely mechanical variable to tune both the electron bandwidth and the structural transition pathway. This makes buckled SiKL an appealing candidate for exploring interaction-driven magnetism, superconductivity, and topological phenomena under strain, doping, or electrostatic gating, while retaining a credible connection to silicon epitaxy and microelectronic processing.

Several targeted calculations can now build on this foundation: resolving the residual phonon anharmonicity; evaluating spin-polarized competing states, orbital projections, full-Brillouin-zone bandwidths, and interaction scales; unfolding finite-temperature spectra; and determining Ag adhesion, registry, reconstruction, and kinetics with a first-principles or validated reactive model. While these studies can sharpen the quantitative phase boundaries and synthesis window, and are the scope of future work, the central qualitative results are already supported across the hierarchy of calculations used here.
\appendix
\section{Planar group-14 kagome-graphene analogues}
\label{app:group14}

For comparison, planar networks with the same bond-linked triangular topology were relaxed for C, Si, Ge, Sn, and Pb using PBE and LDA. In each case, geometry optimization preserves the imposed planar connectivity, and the calculated bands contain weakly dispersive states and sharp DOS features associated with the common kagome-derived interference pattern. Geometry relaxation alone does not establish dynamical stability, so phonon calculations and broader structural searches remain necessary for the heavier analogues.

\begin{figure}
\centering
\includegraphics[width=0.80\textwidth]{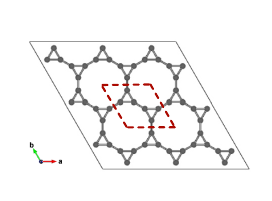}
\caption{Top view of a $3\times3$ supercell of the planar bond-linked group-14 network. The dashed polygon marks one primitive cell.}
\label{fig:group14_structure}
\end{figure}

\begin{table}[h]
\caption{Relaxed lattice constant $a$, intra-triangle bond length $d_1$, inter-triangle bond length $d_2$, and pore diameter $D_p$ for planar group-14 analogues.}
\label{tab:group14}
\centering
\footnotesize
\renewcommand{\arraystretch}{1.15}
\begin{tabular}{|l|l|c|c|c|c|}
\hline
Structure & Functional & $a$ (\AA) & $d_1$ (\AA) & $d_2$ (\AA) & $D_p$ (\AA) \\
\hline
CKL  & LDA & 5.1522  & 1.3441 & 1.4121 & 5.1934 \\
CKL  & PBE & 5.1823  & 1.3545 & 1.4240 & 5.2334 \\
SiKL & LDA & 8.2920  & 2.1985 & 2.2420 & 8.4943 \\
SiKL & PBE & 8.3826  & 2.2238 & 2.2654 & 8.5922 \\
GeKL & LDA & 8.8539  & 2.3345 & 2.4053 & 9.0197 \\
GeKL & PBE & 9.1351  & 2.4258 & 2.4667 & 9.3727 \\
SnKL & LDA & 10.1841 & 2.7827 & 2.8091 & 10.7513 \\
SnKL & PBE & 10.5985 & 2.7833 & 2.8094 & 10.7537 \\
PbKL & LDA & 10.8659 & 2.9612 & 2.9455 & 11.3798 \\
PbKL & PBE & 11.3080 & 3.1079 & 3.0600 & 12.1456 \\
\hline
\end{tabular}
\end{table}

\begin{figure}
\centering
\includegraphics[width=1.0\textwidth]{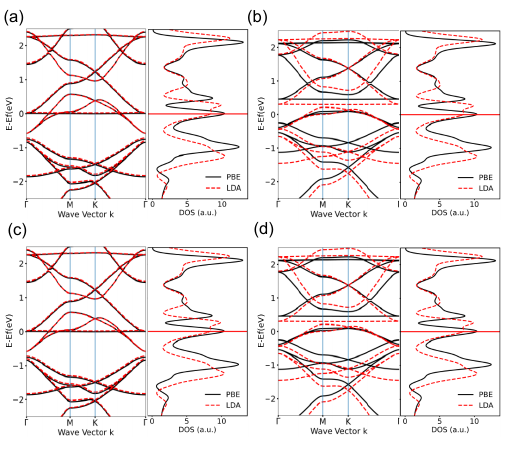}
\caption{Band structures (left) and densities of states (right) of planar (a) Si, (b) Ge, (c) Sn, and (d) Pb linked-triangle lattices calculated with LDA (red dashed curves) and PBE (black solid curves).}
\label{fig:group14_bands}
\end{figure}

Figure~\ref{fig:group14_bands} shows the PBE and LDA electronic structures of the Si, Ge, Sn, and Pb analogues. The common topology produces a flat or weakly dispersive band in each case, while increasing orbital complexity introduces additional hybridization and crossings. In the Sn lattice, the weakly dispersive state lies approximately 0.5~eV above the Fermi level, suggesting that carrier doping or electrostatic gating could be used to approach it; a quantitative carrier-density estimate is left for future work.

\section*{Acknowledgements}
This work was supported by grant DE-SC0023432 funded by the U.S. Department of Energy, Office of Science. This research used resources of the National Energy Research Scientific Computing Center, a DOE Office of Science User Facility supported by the Office of Science of the U.S. Department of Energy under Contract No.~DE-AC02-05CH11231, using NERSC awards BES-ERCAP0037205, BES-ERCAP0033206, BES-ERCAP0025205, BES-ERCAP0025168, and BES-ERCAP0028072. ASB acknowledges startup support from the UCLA Samueli School of Engineering and funding from UCLA's Council on Research Faculty Research Grant. The authors acknowledge the use of the GPT-5 (OpenAI) model to polish the language and edit grammatical errors in some sections of this manuscript. The authors subsequently inspected, validated and edited the text generated by the AI model, before incorporation.

\section*{Data availability}
The data that support the findings of this study are available from the corresponding author upon reasonable request.

\section*{References}
\bibliographystyle{iopart-num}
\bibliography{main}

\end{document}